\documentclass{pasj00}
\draft

\begin{document}
\SetRunningHead{W.-S. Jeong et al.}{The Far-Infrared Properties of Spatially
Resolved AKARI Observations}
\Received{2007/02/25}
\Accepted{2007/00/00}

\title{The Far-Infrared Properties of Spatially Resolved AKARI Observations}

\author{Woong-Seob \textsc{Jeong}\altaffilmark{1},
Takao \textsc{Nakagawa}\altaffilmark{1}, Issei
\textsc{Yamamura}\altaffilmark{1}, Chris P. \textsc{Pearson}\altaffilmark{1,2},
Richard S. \textsc{Savage}\altaffilmark{3}, Hyung Mok
\textsc{Lee}\altaffilmark{4}, Hiroshi \textsc{Shibai}\altaffilmark{5},
Sin'itirou \textsc{Makiuti}\altaffilmark{1}, Hajime
\textsc{Baba}\altaffilmark{6} Peter \textsc{Barthel}\altaffilmark{7} Dave
\textsc{Clements}\altaffilmark{8} Yasuo \textsc{Doi}\altaffilmark{1} Elysandra
\textsc{Figueredo}\altaffilmark{9} Tomotsugu \textsc{Goto}\altaffilmark{1}
Sunao \textsc{Hasegawa}\altaffilmark{1} Hidehiro
\textsc{Kaneda}\altaffilmark{1} Mitsunobu \textsc{Kawada}\altaffilmark{5} Akiko
\textsc{Kawamura}\altaffilmark{5} Do \textsc{Kester}\altaffilmark{10} Suk Minn
\textsc{Kwon}\altaffilmark{11} Hideo \textsc{Matsuhara}\altaffilmark{1} Shuji
\textsc{Matsuura}\altaffilmark{1} Hiroshi \textsc{Murakami}\altaffilmark{1}
Sang Hoon \textsc{Oh}\altaffilmark{4} Sebastian \textsc{Oliver}\altaffilmark{3}
Soojong \textsc{Pak}\altaffilmark{12} Yong-Sun \textsc{Park}\altaffilmark{4}
Michael \textsc{Rowan-Robinson}\altaffilmark{8} Stephen
\textsc{Serjeant}\altaffilmark{9} Mai \textsc{Shirahata}\altaffilmark{1}
Jungjoo \textsc{Sohn}\altaffilmark{4} Toshinobu \textsc{Takagi}\altaffilmark{1}
Lingyu \textsc{Wang}\altaffilmark{8} Glenn J. \textsc{White}\altaffilmark{9,13}
Chisato \textsc{Yamauchi}\altaffilmark{1}}

\altaffiltext{1}{Institute of Space and Astronautical Science, Japan Aerospace
Exploration Agency, Yoshinodai 3-1-1, Sagamihara, Kanagawa 229-8510, Japan}
\email{jeongws@ir.isas.jaxa.jp}

\altaffiltext{2}{European Space Astronomy Centre (ESAC), Apartado 50727, 28080
Madrid, Spain}

\altaffiltext{3}{Astronomy Centre, Department of Physics and Astronomy,
University of Sussex, Brighton BN1 9QH, UK}

\altaffiltext{4}{Department of Physics and Astronomy, Seoul National
University, Shillim-dong, Kwanak-gu, Seoul 151-742, South Korea}

\altaffiltext{5}{Infrared Astrophysics Laboratory, Nagoya University, Furo-cho,
Chikusa-ku, Nagoya 464-8602, Japan}

\altaffiltext{6}{Center for Research and Development, University of Education,
Ibaraki University, Japan}

\altaffiltext{7}{Kapteyn Astronomical Institute, University of Groningen, The
Netherlands}

\altaffiltext{8}{Astrophysics Group,Imperial College of Science Technology and
Medicine, UK}

\altaffiltext{9}{Astrophysics Group, Department of Physics, The Open
University, UK}

\altaffiltext{10}{Netherlands Institute for Space Research SRON, Groningen, The
Netherlands}

\altaffiltext{11}{Department of Science Education, Kangwon National University,
Hyoja-dong, Chunchon-si, Kangwon-do 200-701, South Korea}

\altaffiltext{12}{Department of Astronomy and Space Science, Kyung Hee
University, 1 Seocheon-dong, Giheung-gu, Yongin-si, Gyeonggi-do 446-701, South
Korea}

\altaffiltext{13}{Space Science \& Technology Department, CCLRC Rutherford,
Appleton Laboratory, Chilton, Didcot, Oxfordshire OX11 0QX, UK}


%


\maketitle

\begin{abstract}
We present the spatially resolved observations of IRAS sources from the
Japanese infrared astronomy satellite AKARI All-Sky Survey during the
performance verification (PV) phase of the mission. We extracted reliable point
sources matched with IRAS point source catalogue. By comparing IRAS and AKARI
fluxes, we found that the flux measurements of some IRAS sources could have
been over or underestimated and affected by the local background rather than
the global background. We also found possible candidates for new AKARI sources
and confirmed that AKARI observations resolved IRAS sources into multiple
sources. All-Sky Survey observations are expected to verify the accuracies of
IRAS flux measurements and to find new extragalactic point sources.
\end{abstract}

\section{Introduction}\label{sec:intro}

The Infrared Astronomy Satellite mission (IRAS) \citep{soifer87} successfully
performed the first all-sky survey at infrared wavelengths, providing many
fruitful results. After this pioneering  mission over two decades ago, the
AKARI satellite (previously known as ASTRO-F)
\citep{mura98,naka01,shib04,cpp04,murakami07} is performing the next generation
all-sky survey primarily with the FIS (Far-Infrared Surveyor) focal plane
instrument in 4 far-IR bands (50--200 $\mu$m range) to improved spatial
resolutions and wider wavelength coverage than the its predecessor IRAS [see
\citet{kawada07} for detailed specifications and hardware performance of the
FIS instrument].

Although the telescope apertures of IRAS (60 cm) and AKARI (68.5 cm) are
similar, the effective beam size of the AKARI mission (0.5$^\prime$ $-$
0.9$^\prime$) is much smaller than that of IRAS (2$^\prime$ $-$ 5$^\prime$)
because the sizes of the detector pixels in IRAS were much larger than the size
of the diffracted beam of the telescope. Due to this relatively large beam
size, the astronomical observations made in the far-IR bands of  the IRAS
mission were presumed to have been affected by the fluctuation noise arising
from both background structures and many overlapping extragalactic point
sources below the resolution of the beam. As the resolution of the
telescope/instrument beam increases, the measurements of point source fluxes
are less prone to the fluctuation noises. Thus, fluctuation noise needs to be
more carefully treated for accurate extraction of sources and flux measurements
[see Jeong et al. (2005, 2006) for realistic estimations of these noises for
the AKARI mission]. From observations of sources for those with IRAS fluxes
made during the Performance Verification (PV) phase of the AKARI, we
investigate how the resolved background near a point source can affect the
measurement of fluxes by comparing the measured fluxes.

This paper is organized as follows. In Section \ref{sec:ass_akari}, we describe
the data and the data reduction. We present the comparison of our results
between IRAS flux and AKARI flux in Section \ref{sec:comp_flux}. The probable
detections of new AKARI sources are explained in Section \ref{sec:new_akari}.
We summarize our results in Section \ref{sec:summary}.


\section{AKARI All-Sky Observation and Data Reduction}
\label{sec:ass_akari}

\begin{figure*}
  \begin{center}
    \FigureFile(90mm, 80mm){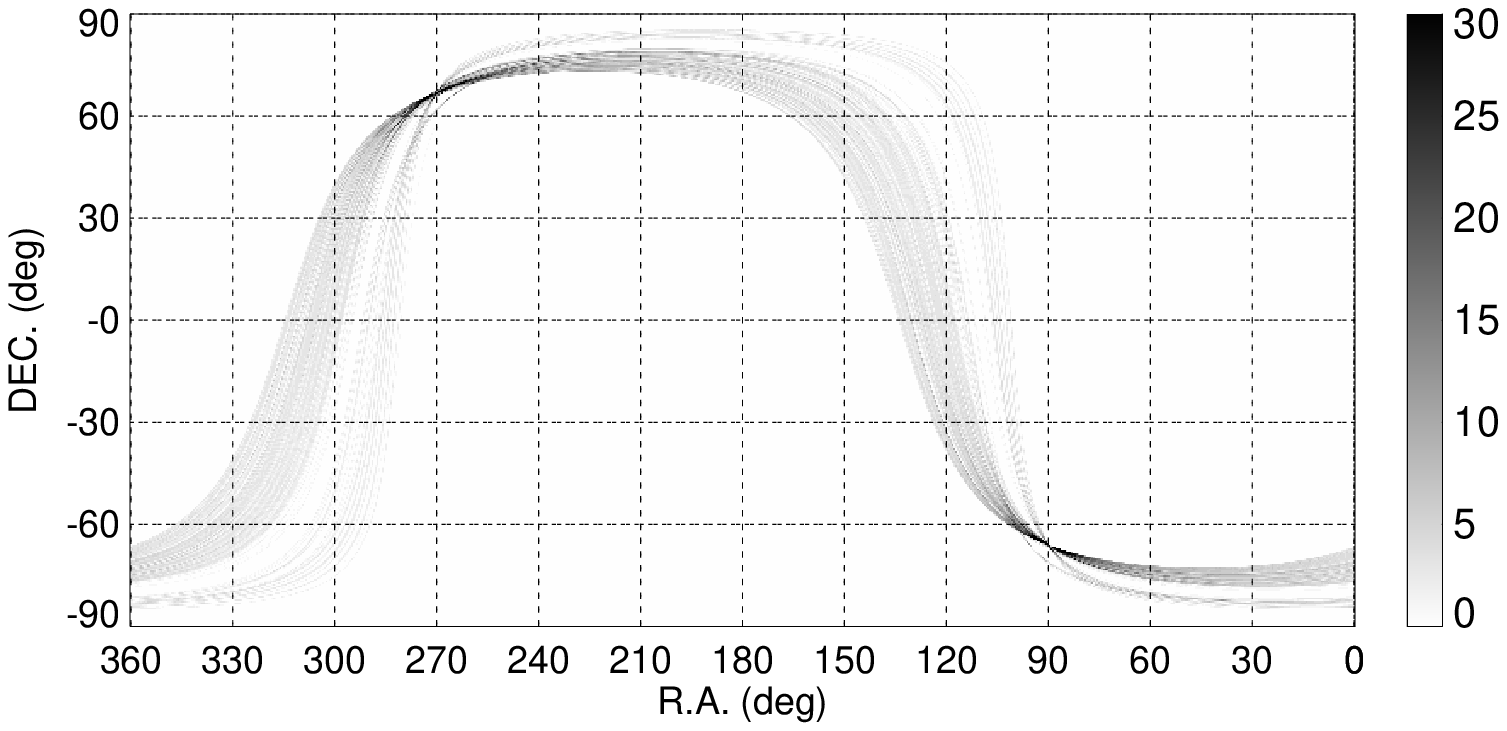}
    \hspace{-0.5in}
    \FigureFile(90mm, 80mm){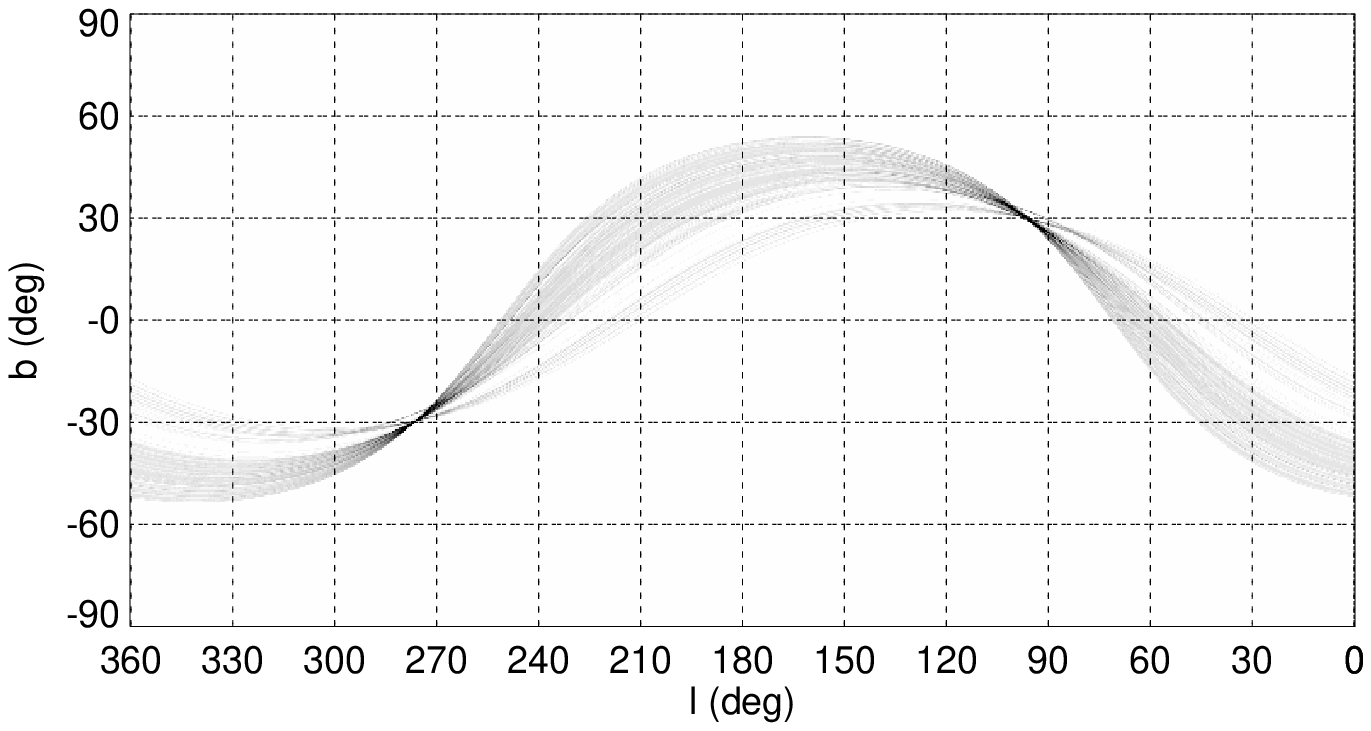}
    \vspace{-0.3in}
  \end{center}
  \caption{Scan map area during the Performance Verification (PV) phase in
  Equatorial coordinates (left) and Galactic coordinates (right). The gray bar
  shows the number of orbit surveyed by AKARI. Due to the nature of AKARIs
  orbit, the regions near the North Ecliptic Pole (NEP) and the South Ecliptic
  Pole (SEP) have visibility significantly higher than other regions on the
  celestial sphere.}
  \label{fig:scanmap_pv}
\end{figure*}

The data used in this work are from AKARI All-Sky Survey observations performed
during the two week-long PV phase from April 24th 2006 to May 7th 2006. The
observations cover a range in Galactic latitude between -50$^\circ$ to
50$^\circ$ as shown in figure \ref{fig:scanmap_pv}.

\begin{figure*}
  \begin{center}
    \FigureFile(90mm, 90mm){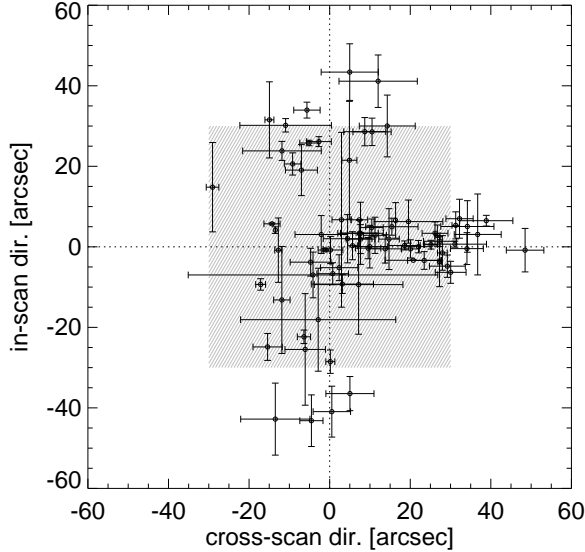}
    \vspace{0.4in}
  \end{center}
  \caption{Positional difference between IRAS and AKARI for samples with
  a good IRAS flux quality. The shaded area means the pixel size of AKARI/FIS
  for the WIDE-S band (30$''$). Since we only plot sources confirmed in at
  least two more orbit, the error bar means the standard deviation of
  positions in in-scan and cross-scan direction.}
  \label{fig:pos_diff_det}
\end{figure*}

\begin{figure*}
  \begin{center}
    \FigureFile(130mm, 80mm){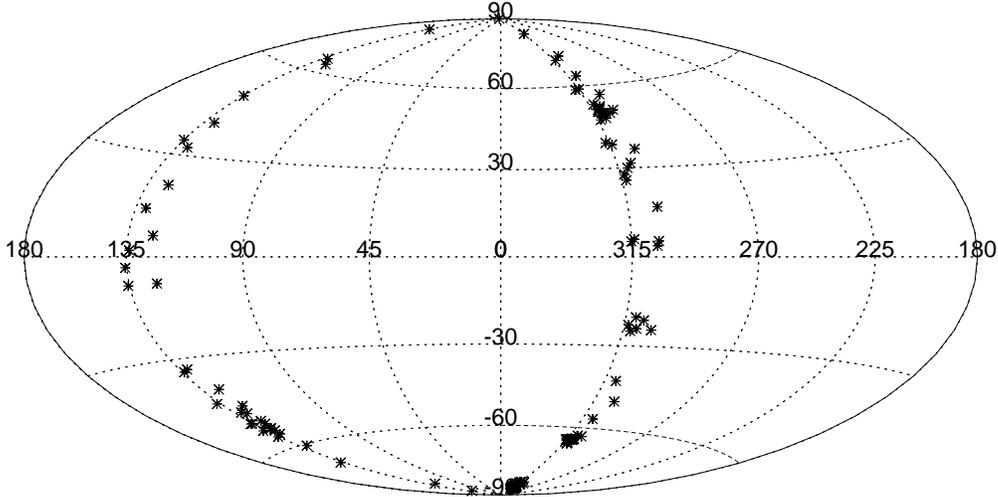}
    \vspace{-0.4in}
  \end{center}
  \caption{Distribution of sources matched with IRAS catalogue
  in ecliptic coordinates.}
  \label{fig:pos_det_el}
\end{figure*}

For the data processing, we have used the latest pipeline developed for the
All-Sky Survey \citep{kawamura03}. The source extraction was performed only for
the AKARI WIDE-S band (90$\mu$m) with the dedicated software for the AKARI
survey. On the purpose of calibrating the absolute flux, the observations of
asteroids, stars and galaxies with a wide range of fluxes were used, and the
uncertainty of flux calibration is expected to be less than 20\% for WIDE-S
band \citep{kawada07}. In order to avoid false detections, we removed any
detections near the South Atlantic Anomaly (SAA) region, reset or calibration
sequences and the edge of a detector array. Among the extracted sources, we
have selected reliable sources with the following criteria to eliminate false
detections: (1) signal to noise ratio (S/N) greater than 15, (2) detected in
all rows of the detector array, e.g., 3 rows of detector pixels for the WIDE-S
band, and (3) confirmed in at least two more orbit ($\sim$ 100 min. per 1 orbit
for AKARI) (4) the Galactic plane region is excluded as it is observed in a
special detector readout mode that needs separate processing [for the sampling
mode of AKARI observation, see \citet{kawada07} for detailed information].
After selecting reliable sources, we attempted to match them with IRAS sources
within a positional error circle of 100$''$. The total number of sources in our
final selected sample is 150. We showed the positional difference between AKARI
and IRAS sources with a good flux quality in Figure \ref{fig:pos_diff_det}.
Most of positions for IRAS sources are well consistent with those estimated in
AKARI within a resolution range (30$''$). The current position information is
determined based on the sensors of the onboard attitude and orbit control
system (AOCS), reprocessed on the ground. The major source of uncertainty is
the alignment between the AOCS sensors and the telescope, and its time
variation. In the future, position information will be improved with the
pointing reconstruction processing using the data from the focal-plane
instruments. The accuracy of the detected source position will be as good as a
few arcsec, which enable us to carry out more detailed analysis of the `IRAS'
positions. Figure \ref{fig:pos_det_el} shows the distribution of these sources
in ecliptic coordinates.

\section{A Comparative Study of AKARI and IRAS Fluxes}
\label{sec:comp_flux}

The reliably detected sources are used to compare the IRAS and AKARI fluxes.

\subsection{Flux Comparison}

\begin{figure}
  \begin{center}
    \FigureFile(130mm, 130mm){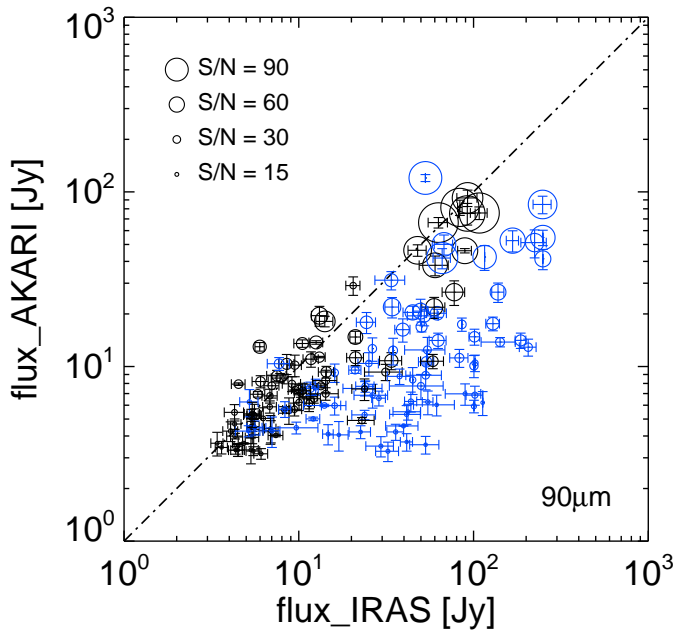}
  \end{center}
  \caption{Comparison of flux between IRAS and AKARI.
  The larger symbol means the detection with larger S/N. The black symbols
  is for IRAS sources with high flux quality both in 60 and 100$\mu$m and
  the other cases are displayed in the blue symbols. We also plot the error
  bars for AKARI and IRAS flux.}
  \label{fig:comp_flux}
\end{figure}

The effective wavelength of the AKARI WIDE-S band is 90$\mu$m while IRAS has
60$\mu$m and 100$\mu$m bands in the far-IR region. In order to make a
consistent comparison between the AKARI and IRAS fluxes, we computed the IRAS
flux at 90$\mu$m by fitting the 60 and 100 $\mu$m data to a black-body curve.
Since the flux densities of IRAS point sources listed in the catalogue are
non-color corrected ones, we attempted an iterative fitting method which
performs successively fitting and color corrections. We also applied the color
correction to the AKARI flux densities. Figure \ref{fig:comp_flux} shows the
comparison between the AKARI and IRAS fluxes of our detected sources at
90$\mu$m. We also plot the error bars for the IRAS and AKARI fluxes. Note that
a flux error for an AKARI flux was obtained from the standard deviation of the
extracted fluxes for orbit-confirmed detections while the flux error for the
IRAS sources with a bad flux quality was not listed in IRAS point source
catalogue. The straight line shows the perfect correlation between these two
fluxes. The size of the circle is proportional to the estimated signal-to-noise
ratio. This figure shows that the data lies below the envelope of
$F_{AKARI}\approx F_{IRAS}$. This means that the AKARI fluxes are in general
similar to or smaller than the IRAS fluxes.

\begin{figure*}
  \begin{center}
    \FigureFile(120mm, 120mm){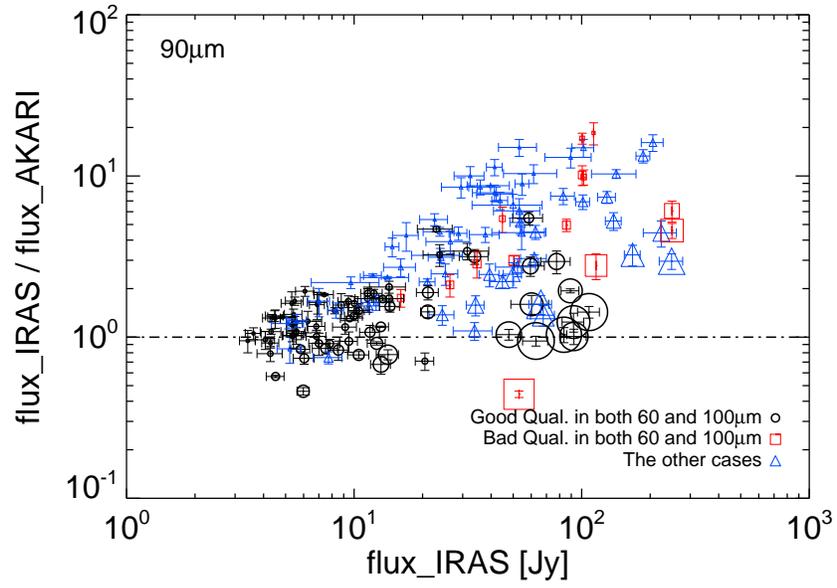}
    \FigureFile(120mm, 120mm){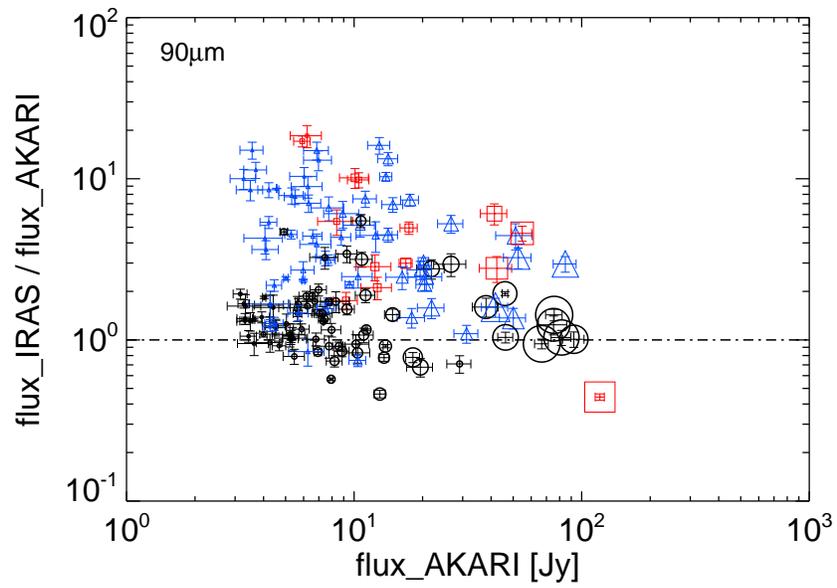}
  \end{center}
  \caption{Flux ratios of IRAS to AKARI flux measurements
  as a function of IRAS flux (upper) and AKARI flux (lower).}
  \label{fig:comp_flux_ratio}
\end{figure*}

Such a trend is more clearly shown in figure \ref{fig:comp_flux_ratio} which
displays the flux ratio of IRAS to AKARI detections. We checked the quality of
the IRAS flux, by examining the quality flag for the deviating sources. Most of
those sources indeed have a moderate or a bad flux quality and are often
located near the Galactic plane or strong star formation regions (e.g., Large
Magellanic Cloud). Thus, due to a strong background, the flux of IRAS sources
with moderate or bad flux quality flags may not have been accurately estimated.
We also found that the flux is over or underestimated even for IRAS sources
with a high flux quality. The scattering trend in flux ratio shows a triangular
shape (peaking at 80 Jy) in the upper panel of figure \ref{fig:comp_flux_ratio}
and a right triangular shape in the lower panel of figure
\ref{fig:comp_flux_ratio}, respectively. The sensitivity of AKARI in the low
flux range is better than that of IRAS, which explains that the scatter in the
lower panel of figure \ref{fig:comp_flux_ratio} is larger for low fluxes than
for high fluxes.

\subsection{Effect of the Background on the Flux Estimation}\label{subsec:cont_bgr}

\begin{figure*}
  \begin{center}
    \FigureFile(120mm, 120mm){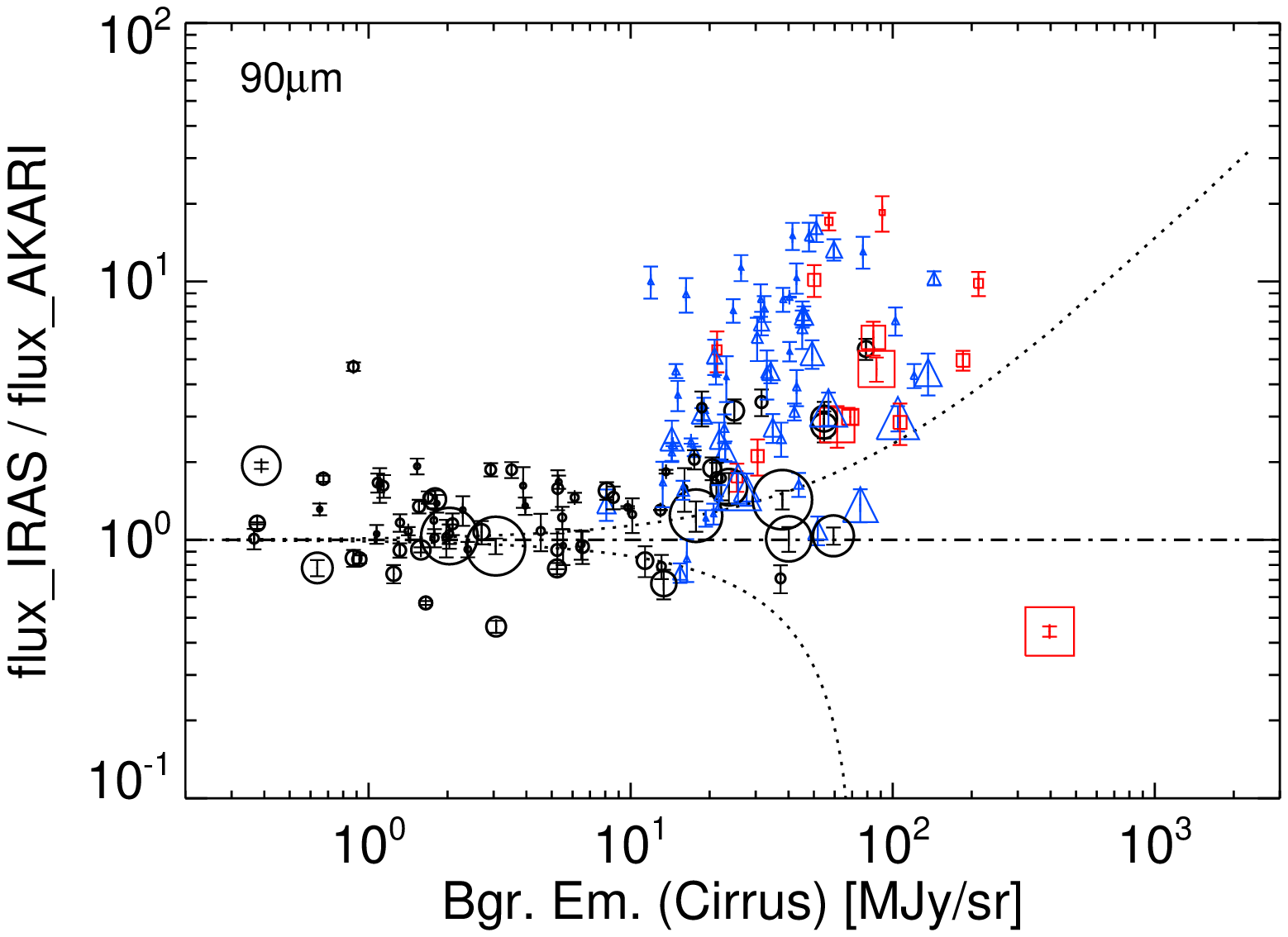}
    \FigureFile(120mm, 120mm){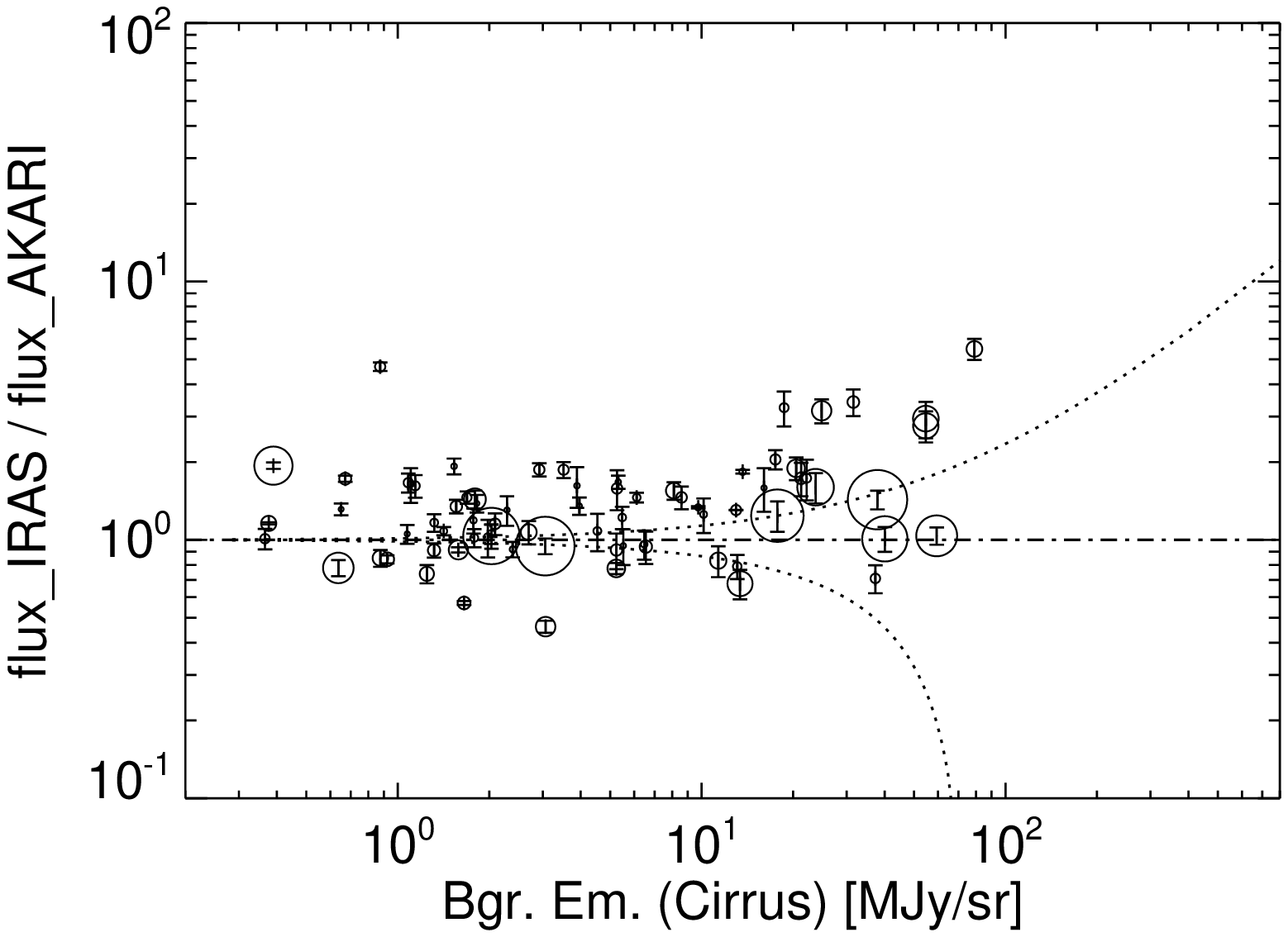}
  \end{center}
  \caption{Ratio of IRAS to AKARI flux measurements.
  We plot all matched sources (upper) and sources matched with IRAS
  sources with a high quality flag (lower). The dotted line shows
  the expected contamination from the background for 10 Jy source.
  The explanations for symbols are the same in the caption of figure
  \ref{fig:comp_flux_ratio}.}
  \label{fig:comp_flux_bgr}
\end{figure*}

The background fluctuations from the surface brightness of extended structure,
can easily be mistaken for genuine point sources. Since our detections are for
bright sources, the contribution of the fluctuation noise will be primarily
from sky confusion (rather than point sources). Some authors have estimated the
sky confusion from background fluctuations for the AKARI mission based upon
IRAS and ISO data \citep{jeong05,kiss05}. In order to check any dependency of
the flux uncertainty on the background level, we plot flux ratios as a function
of the background emission in figure \ref{fig:comp_flux_bgr}. We assume that
the fluctuations from the background are mainly from cirrus emission. The
fluctuations from the zodiacal light is expected to be small at the resolution
of the IRAS mission. For the estimation of the cirrus emission in the AKARI
band, we used the all-sky 100 $\mu$m dust map generated from the IRAS and COBE
data by Schlegel, Finkbeiner, and Davis (1998) and the dust model by Finkbeiner
et al. (1999). As shown in the upper panel of figure \ref{fig:comp_flux_bgr},
the scatter increases around the 20 MJy/sr level, with an underlying monotonic
increase from 20 to 200 MJy/sr. Owing to its higher resolution, AKARI can
resolve the background more effectively. Since the estimated background for
IRAS depends on the relatively larger spatial structure of the background
compared to the diffracted beam size [the source extraction for IRAS used an
annulus of radius 5$'$$-$7$'$ in measuring the background \citep{wheel94}], it
could not be measured properly on a intensively fluctuating background.
Therefore, this trend shows that the local background at least below 5$'$ near
a source, rather than the global background, is an important factor in the flux
estimation. Note that previous estimations of sky confusion
\citep{jeong05,kiss05} suggested that the fluctuation noise is strongly
dependent on the background level with the assumption that the spatial power
spectrum from fluctuations of the dust emission has a simple power-law with the
power index of $\sim$-3 \citep{gautier92}. In addition, we have also made a
simple estimation of the contamination of the flux by summing the background
flux over a detector pixel. We plot the case for a 10 Jy IRAS source
contaminated by a 10 percent background (see the dotted line in figure
\ref{fig:comp_flux_bgr}). We found that the flux measurements for some IRAS
sources might be contaminated by the background. If we plot AKARI sources
matched with IRAS sources with only high flux quality flags (see the lower
panel of figure \ref{fig:comp_flux_bgr}), the scatter is reduced, however,
visible scatter still remains and slightly increases as the cirrus background
becomes larger. Thus, IRAS flux measurements may have errors even for correctly
measured IRAS sources with a high flux quality. For a more systematic study to
confirm this result, we need to use samples from a wider survey area.

\section{New AKARI Sources}\label{sec:new_akari}

\begin{figure*}
  \begin{center}
    \FigureFile(70mm, 70mm){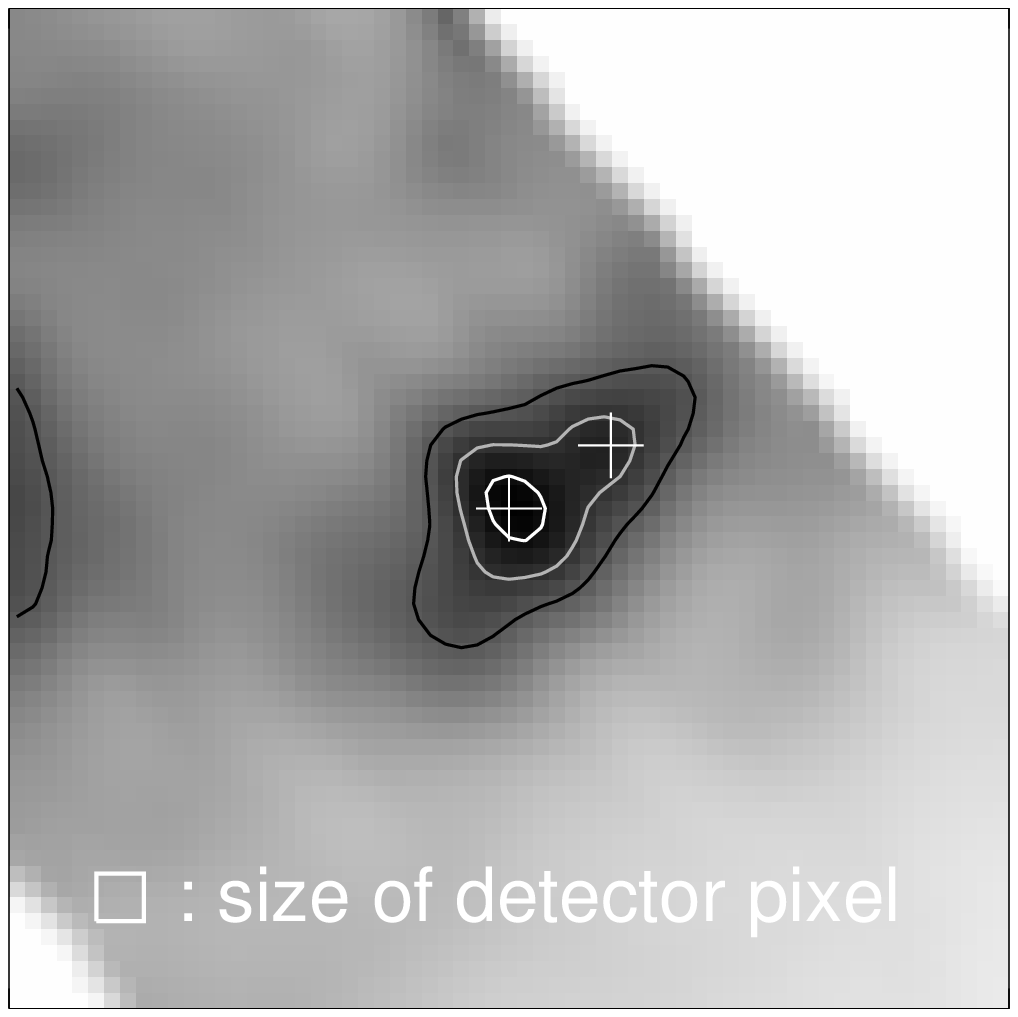}
    \hspace{-0.9in}
    \FigureFile(70mm, 70mm){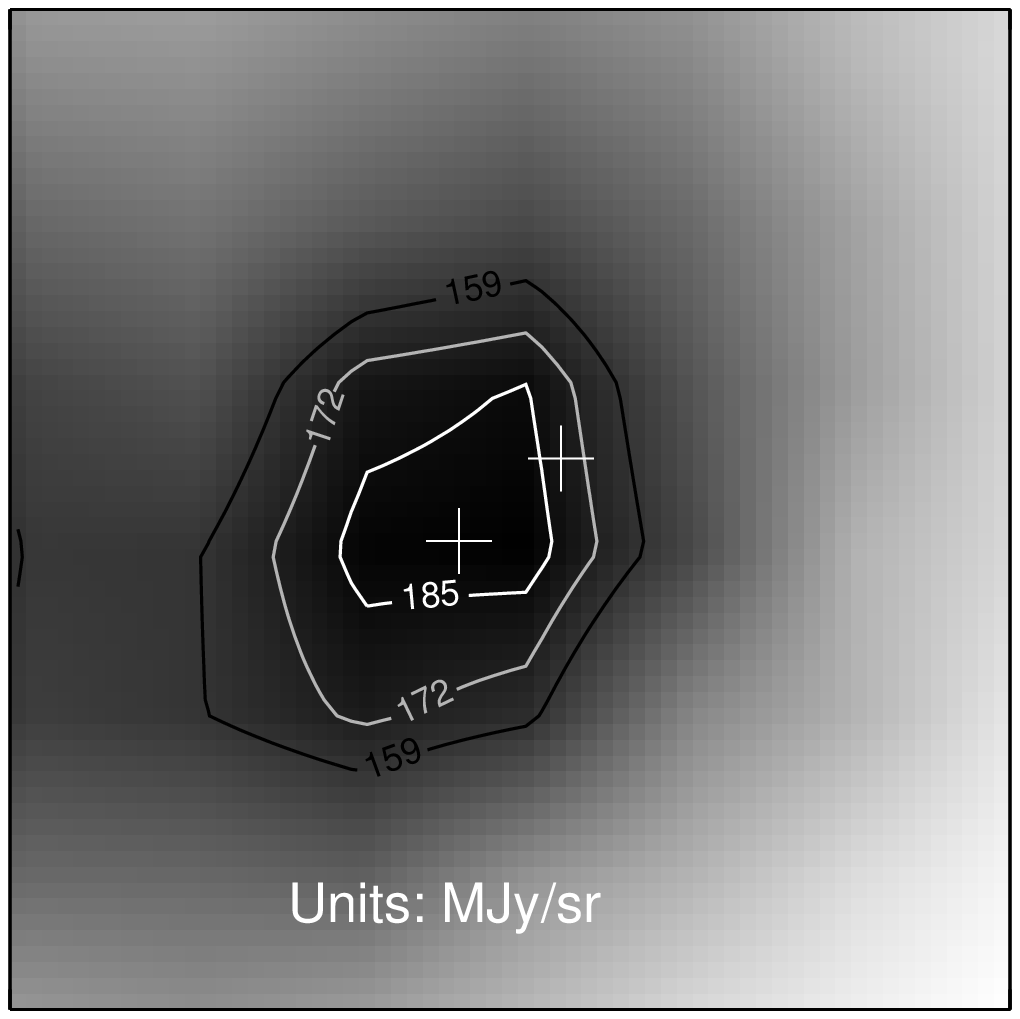}
    \hspace{-0.9in}
    \FigureFile(70mm, 70mm){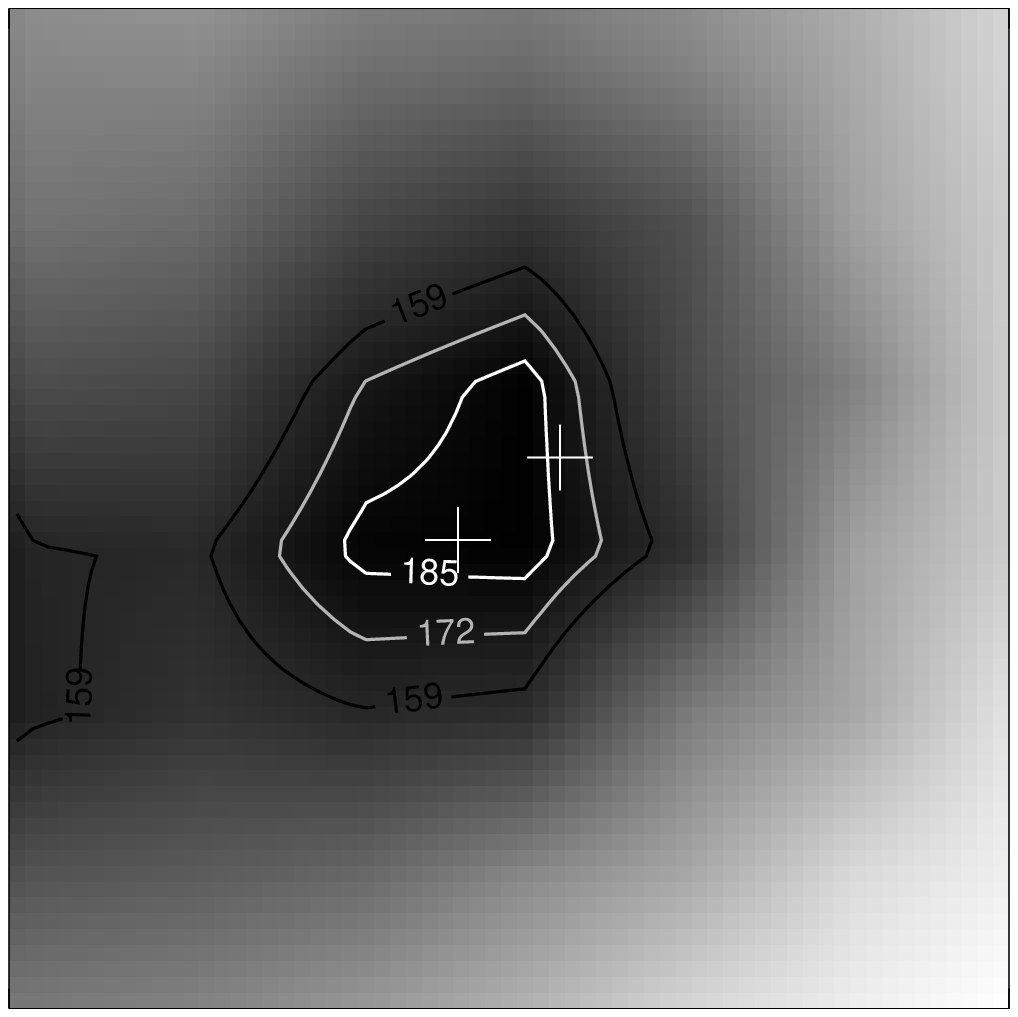}
  \end{center}
  \caption{Detection image for IRAS05360-6914 with the size of
  10.5$'$ $\times$ 10.5$'$ from AKARI 90$\mu$m (left), IRAS 60$\mu$m
  (center), and 100$\mu$m (right) map in linear gray scale for flux.
  We used the revised IRAS map by \citet{miville05} which included
  well-calibrated point sources and corrected diffuse emission
  calibration. Since AKARI image is not calibrated yet, we did not show
  contour levels. The flux for IRAS source listed in the catalogue
  is 61.67 (60$\mu$m) and 251.8 Jy (100$\mu$m), and the flux quality
  was measured to be bad (60$\mu$m) and moderate (100$\mu$m). Owing to
  the higher resolution of AKARI, this IRAS source is detected as two
  sources in AKARI observation. In addition, we found that the flux
  estimated from IRAS is also overestimated in this detection by
  a factor of $\sim$5. The plus symbols mean the positions of two
  detected sources.}
  \label{fig:det_new_akari}
\end{figure*}

In this work, we have used only sources matched with IRAS point source
catalogue. However, we also detected new sources from the AKARI All-Sky Survey
which are not present in the IRAS catalogue. There are many possible candidates
for these new AKARI sources. First, one IRAS source may have been resolved into
two sources (see one example shown in figure \ref{fig:det_new_akari}). Second,
some sources may be resolved from the background. Third, the source may be a
Solar System object, e.g., asteroid. Fourth, some detections may be false due
to the detector characteristics. On the other hand, there may be false
detections from IRAS. Thus, we expect that the AKARI All-Sky Survey can find
new extragalactic point sources as well as faint galactic point sources.

\section{Summary}\label{sec:summary}

By using the AKARI All-Sky Survey data taken during the PV phase, we tried to
extract reliably detected sources and compared their fluxes with associations
in the IRAS point source catalogue. The positions of IRAS sources are mostly
well consistent with those estimated from AKARI. We found that the original
flux measurements of some IRAS sources appear to have been overestimated or
underestimated. We also confirmed that most of the IRAS sources with a moderate
or a bad quality flux flag in the IRAS catalogue were quite overestimated. To
investigate any possible contamination of the IRAS source flux from the
background in the estimation of the flux, we checked the background level on
each detection of the IRAS sources. The difference in the flux between the IRAS
and AKARI associations was not found to be strongly dependent on the background
level. However, it is suggested that the flux measurement may have been
affected by the local background rather than the global background. Note that
AKARI can resolve the background and source more effectively than IRAS due to
the higher resolution of the AKARI observations. In addition, we found that in
some cases, the flux measurements even for correctly measured (high flux
quality) IRAS sources may include errors of up to $\sim$5, even though we
consider the uncertainty of flux calibration. However, due to various detector
characteristics, there may be false detections as well as detections with
inaccurately estimated flux though we have removed false detections and
selected reliable detections as described in Section \ref{sec:ass_akari}. The
AKARI All-Sky Survey is expected both to verify the IRAS sources and find new
extragalactic or galactic point sources. The preliminary catalogue for reliably
detected point sources will be released in a forthcoming paper.

\section*{Acknowledgments}

The AKARI Project is an infrared mission of the Japan Space Exploration Agency
(JAXA) Institute of Space and Astronautical Science (ISAS), and is carried out
with the participation of mainly the following institutes; Nagoya University,
The University of Tokyo, National Astronomical Observatory Japan, The European
Space Agency (ESA), Imperial College London, University of Sussex, The Open
University (UK), University of Groningen / SRON (The Netherlands), Seoul
National University (Korea). The far-infrared detectors were developed under
collaboration with The National Institute of Information and Communications
Technology. We would like to thank the referee L. Viktor T$\acute{\rm o}$th for
a careful reading of our paper and many fruitful comments. W.-S. Jeong
acknowledges a Japan Society for the Promotion of Science (JSPS) fellowship to
Japan. We thank Myungshin Im and Bon-Chul Koo for helpful suggestions.



\end{document}